\documentclass[twocolumn]{jpsj2}
\usepackage{graphicx} % Include figure files
\usepackage[dvipdfm]{color}
\usepackage{amsmath}
\usepackage{txfonts}
\usepackage{bm}

\newcommand{\sub}[1]{$_{\mathrm {#1}}$}
\newcommand{\subm}[1]{_{\mathrm {#1}}}

\newcommand{\spsm}[1]{^{\mathrm {#1}}}

\newcommand{\etal}{\textit{et~al.}}

\renewcommand{\deg}{^{\circ}}
\newcommand{\Tc}{T\subm{c}}
\newcommand{\Tco}{T\subm{c}\spsm{onset}}
\newcommand{\Tcj}{T\subm{c}\spsm{junction}}
\newcommand{\Tcjz}{T\subm{c0}\spsm{junction}}
\newcommand{\Tcz}{T\subm{c0}}
\newcommand{\Hvec}{{\boldmath$H$}}
\newcommand{\Hbm}{\bm{H}}
\newcommand{\x}{{\boldmath$X$}}
\newcommand{\Xbm}{\bm{X}}

\newcommand{\Hcc}{H\subm{c2}}

\newcommand{\dash}{^{\prime}}
\newcommand{\tm}[1]{(TMTSF)\sub{2}{#1}}
\newcommand{\tmc}{\tm{ClO\sub{4}}}
\newcommand{\tmp}{\tm{PF\sub{6}}}
\newcommand{\tmx}{\tm{$X$}}
\newcommand{\cstar}{c^{\ast}}
\newcommand{\Rzz}{R_{c^{\ast}}}
\newcommand{\Hp}{H\subm{P}}
\newcommand{\Horb}{H\subm{c2}^{\mathrm{orb}}}
\newcommand{\qFFLO}{\bm{q}_{\mathrm{FFLO}}}
\newcommand{\qnest}{\bm{q}_{\uparrow\downarrow}}

\newcommand{\kbm}{\bm{k}}

\newcommand{\phidco}{\phi\subm{3D-2D}}
\newcommand{\phidip}{\phi\subm{dip}}
\newcommand{\ta}{t_a}
\newcommand{\tb}{t_{b\dash}}
\newcommand{\tc}{t_{\cstar}}

\title{%
Magnetic-Field Variations of the Pair-Breaking Effects of Superconductivity in (TMTSF)$_2$ClO$_4$
}

\author{%
Shingo~Yonezawa\thanks{E-mail address: yonezawa@scphys.kyoto-u.ac.jp},
Shuichi~Kusaba, Yoshiteru~Maeno, Pascale~Auban-Senzier$^{1}$, \\
Claude~Pasquier$^1$, and Denis~J\'{e}rome$^1$
}

\inst{%
Department of Physics, Graduate School of Science, Kyoto University, Kyoto 606-8502, Japan\\
$^2$Laboratoire de Physique des Solides (UMR 8502) - Universit{\'{e}} Paris-Sud, 91405 Orsay, France\\
}

\abst{%
We have studied the onset temperature of the superconductivity $T_{\mathrm{c}}^{\mathrm{onset}}$ 
of the organic superconductor (TMTSF)$_2$ClO$_4$,
by precisely controlling the direction of the magnetic field $\bm{H}$.
We compare the results of two samples with nearly the same onset temperature
but with different scattering relaxation time $\tau$.
We revealed a complicated interplay of a variety of pair-breaking effects and 
mechanisms that overcome these pair-breaking effects.
In low fields, the linear temperature dependences of the onset curves in the $H$-$T$ phase diagrams
are governed by the orbital pair-breaking effect.
The dips in the in-plane field-angle $\phi$ dependence of $T_{\mathrm{c}}^{\mathrm{onset}}$,
which were only observed in the long-$\tau$ sample, provides definitive evidence that
the field-induced dimensional crossover enhances the superconductivity 
if the field direction is more than about 19-degrees away from the $a$ axis.
In the high-field regime for $\bm{H}\parallel a$, 
the upturn of the onset curve for the long-$\tau$ sample indicates a new superconducting state
that overcomes the Pauli pair-breaking effect but is easily suppressed by impurity scatterings.
The Pauli effect is also overcome for $\Hbm\parallel b^{\prime}$ 
by a realization of another state for which the maximum of $\Tco(\phi)$ occurs in a direction 
different from the crystalline axes.
The effect on $T_{\mathrm{c}}^{\mathrm{onset}}$ 
of tilting the applied field out of the conductive plane
suggests that the Pauli effect plays a significant role in determining $T_{\mathrm{c}}^{\mathrm{onset}}$.
The most plausible explanation of these results is 
that (TMTSF)$_2$ClO$_4$ is a singlet superconductor and exhibits Fulde-Ferrell-Larkin-Ovchinnikov 
(FFLO) states in high fields.
}

\kword{%
(TMTSF)$_2$ClO$_4$, Pauli pair-breaking effect, orbital pair-breaking effect, FFLO state, 
field-induced dimensional crossover,
impurity scattering, vortex state
}

\recdate{\today}

\begin{document}

\maketitle

\section{Introduction}
\label{sec:Introduction}

In a spin-singlet type-II superconductor, a magnetic field destroys the superconductivity
mainly through two pair-breaking mechanisms.
One is called the orbital effect, 
which originates from the increase of the kinetic energy of supercurrent around 
the magnetic vortices penetrating the superconductor~\cite{Helfand1966PhysRev}.
When this kinetic energy becomes as large as the condensation energy of the superconductivity,
the superconducting (SC) state becomes unstable.
The other pair-breaking mechanism is the so-called Pauli effect~\cite{Clogston1962},
which results from the Zeeman energy of the quasiparticles.
Assuming the absence of the orbital effect,
the normal state becomes more stable than the SC state
at a magnetic field where normal quasiparticles have lower energy than Cooper pairs
due to the spin polarization.
This characteristic magnetic field is called the Pauli-Clogston limit $\Hp$~\cite{Clogston1962}.
In real materials, the upper critical field $\Hcc$ is determined 
by a combination of these two pair-breaking effects~\cite{Werthamer1966PhysRev}.

However, under certain situations these pair-breaking effects 
become less significant.
For instance, the orbital pair-breaking effect can be suppressed if
the dimensionality of the electronic system is lowered.
In a quasi-one-dimensional (Q1D) or Q2D system,
less supercurrent flows in the less-conductive directions when magnetic fields are applied
parallel to the conductive direction.
As a result, the kinetic energy becomes less in this field configuration,
which makes the orbital effects insignificant.
%In heavy Fermion systems, the orbital effect can be also less important
%because of the slow velocity of the heavy quasiparticles.
The Pauli effect can be negligible if the superconductivity is a spin-triplet pairing state, 
in which carriers form $S=1$ Cooper pairs.
This is because the triplet Cooper pairs can polarize along the field direction 
gaining the Zeeman energy without loosing the condensation energy.
A singlet superconductor can also partly overcome the Pauli effect
by forming a so-called Fulde-Ferrell-Larkin-Ovchinnikov (FFLO or LOFF) state~\cite{Fulde1964,Larkin1965}.
In this state, the carriers on the Zeeman-split Fermi surfaces 
form pairing combinations with a finite momentum of $\qFFLO$.
This finite $\qFFLO$ results in a spatial modulation of the order parameter of superconductivity.
We note that the response of an FFLO state to external parameters 
may differ from the response of the usual SC phase~\cite{Matsuda2007JPhysSocJpnReview}.
For example, an FFLO state is sensitively suppressed 
by a smaller amount of impurities than in the case of usual SC phases~\cite{Gruenberg1966},
because of its broken translational symmetry.
In the case of a 2D system, several authors have suggested that a tilt of the magnetic field
out of the conductive plane leads to an occurrence of unusual vortex states with 
the Landau level indices $n\ge 1$
due to a competition between the spatial structure of the FFLO order parameter
and the structure of the vortex lattice~\cite{Bulaevskii1974JETP,Shimahara1997.JPhysSocJpn.66.3591,Klein2004PhysRevB}.
%These unusual vortex states can be referred to as ``FFLO precursor states'' 
%or ``paramagnetic vortex states''~\cite{Klein2004PhysRevB}.
As we have overviewed, 
there are a number of effects that may break or stabilize superconductivity.
It is quite interesting 
to reveal in detail the interplay of these effects on the stability of superconductivity in magnetic fields.

The organic superconductor \tmx\ 
(where TMTSF stands for tetramethyl-tetraselena-fulvalene and
$X$ is an anion such as ClO\sub{4} or PF\sub{6})~\cite{Jerome1980,Bechgaard1981},
is an ideal material system for studies of the interplay among pair-breaking effects,
the dimensionality of the electronic system, and its superconducting symmetry.
It is an archetypal Q1D system
with a quite anisotropic electronic conductivity.
Therefore, the orbital pair-breaking effect 
becomes insignificant if a magnetic field is applied parallel to the most-conductive $a$ axis.
Moreover, \tmx\ has another mechanism that suppresses the orbital pair-breaking effect:
the field-induced dimensional crossover (FIDC), which was first predicted by Lebed~\cite{Lebed1986}.
If a field is parallel to the $b\dash$ direction, which is perpendicular to the $a$ axis
within the conducting $ab$ plane, 
the temperature dependence of the resistance along the least-conductive $\cstar$ direction 
(the direction perpendicular to the $ab$ plane) $\Rzz(T)$ becomes non-metallic~\cite{Strong1994,Joo2006}.
This behavior can be interpreted as a confinement of carriers in the $ab$ plane.
This phenomenon is called the FIDC, 
because the electronic system become essentially 2D rather than anisotropic 3D.
Therefore, for high magnetic fields,
the orbital pair-breaking mechanism for $\Hbm\parallel b\dash$ can 
become even weaker than for $\Hbm\parallel a$~\cite{Lebed1986}.
We note that, although the FIDC can be interpreted as a result of 
the semi-classical motion of carriers on the sheet-like Fermi surface of \tmx~\cite{Zhang2007AdvPhys},
it is in fact a fully quantum-mechanical phenomenon~\cite{Joo2006,Zhang2007AdvPhys}.

In fact, the $H$-$T$ phase diagram of superconductivity in \tmx\ looks quite unusual.
Lee \etal~\cite{Lee1997} reported that $\Hcc(T)$ of \tmp\ determined from resistivity measurements 
exhibits divergent behavior as temperature decreases 
and $\Hcc(T)$ reaches 80~kOe at the lowest temperature
when magnetic fields are applied parallel to the $b\dash$ axis.
This apparent absence of the orbital pair-breaking in high fields 
is attributable to the FIDC~\cite{Lebed1986}.
An important point is that 
this $\Hcc(T)\parallel b\dash$ does not exhibit any saturation around $\Hp$,
which is estimated to be 20~kOe for their sample,
and that $\Hcc$ at the lowest temperature far exceeds $\Hp$.
Similar results have been obtained also in \tmc\ through resistivity 
and magnetic torque measurements by Oh and Naughton~\cite{Oh2004}
and recently through our resistivity measurements~\cite{Yonezawa2008.PhysRevLett.100.117002}.
It has been speculated that this survival of the superconductivity far above $\Hp$ 
is due to spin-triplet pairing~\cite{Lebed1999,Lebed2000}.
On the other hand, several authors have pointed out that, in a Q1D superconductor, 
even a singlet superconductivity can be stable far above $\Hp$ 
by a realization of an FFLO state~\cite{Dupuis1993,Dupuis1994,Miyazaki1999}, 
because of the nesting nature of its sheet-like Fermi surface.
Recently another possible theoretical scenario,
a transition from a singlet $d$-wave state
to a triplet $f$-wave state at high fields,
has been proposed~\cite{Fuseya2005.JPhysSocJpn.74.1263,Aizawa2007}.
The interpretation of these phase diagrams is still controversial.

In addition, the phase diagram for $\Hbm\parallel a$ also seems unusual.
Near $H=0$, the $\Hcc(T)\parallel a$ curve for \tmp\ 
has a steeper slope at $H=0$ than the $\Hcc(T)\parallel b\dash$ curve,
due to the insignificance of the orbital effect for $\Hbm$ parallel to the most conductive direction.
As temperature decreases, the curve saturates and crosses the $\Hcc(T)\parallel b\dash$ curve 
when it reaches $\Hp\sim 20$~kOe,
and increases again when temperature goes below 0.3~K.
For \tmc, Murata \etal~\cite{Murata1987} reported $\Hcc(T)\parallel a$ above 0.5~K,
which also exhibits a steep initial slope at $H=0$ and a saturation around 30~kOe.
These saturations of $\Hcc(T)\parallel a$ suggest 
that the Pauli effect is important for $\Hbm\parallel a$.
In addition to these features, we revealed in our recent report~\cite{Yonezawa2008.PhysRevLett.100.117002} 
for the first time that
$\Hcc(T)\parallel a$ of \tmc\ exhibits an upturn at lower temperatures,
similar to the behavior in \tmp.

As we have reviewed, there are various effects 
that determine the phase boundary of \tmc\ 
depending on both the field direction and the field amplitude.
Therefore, we are motivated to study the field-amplitude and field-direction dependence 
of the onset temperature of superconductivity.
We note that the in-plane field-angle dependence of $\Hcc$ at 1.03~K of \tmc\ 
was reported~\cite{Murata1987}.
However, at this temperature $\Hcc$ is much lower than $\Hp$ 
and is probably governed by the orbital effect.
We recently reported several anomalous features of the in-plane field-angle dependence of 
the onset temperature $\Tco$ in higher fields~\cite{Yonezawa2008.PhysRevLett.100.117002},
which are attributed to the enhancement of superconductivity due to the FIDC
and a possible realization of FFLO states.

In this work, we study the field-strength dependence and in-plane field-angle dependence of $\Tco$ 
in detail and compare the results of two samples with different qualities, 
in order to reveal the interplay of the orbital effect, the Pauli effect, the FIDC,  
the possible realization of the FFLO state,
and impurity effect.
It is shown that impurity scatterings affect more drastically
the FIDC and the stability of the high-field state for $\Hbm\parallel a$ than 
the stability of the low-field SC state.
Therefore, the dependence on the sample quality of the behavior of $\Tco$ 
reveals not merely the impurity effect on superconductivity,
but provides strong evidence of the enhancement of superconductivity due to the FIDC,
and also information on the difference between the stability against impurity scatterings
of the high-field state and the low-field state.
The change of the behavior of $\Tco$ for fields slightly tilted from the conducting plane
has also been studied,
suggesting a significant contribution of the Pauli effect in determining $\Tco$.

%%%%%%%%%%%%%%%%%%%%%%%%%%%%%%%%%%%%%%%%%Experiments%%%%%%%%%%%%%%%%%%%%%%%%%%%%%%%%%%%%%%%%%%%%
\section{Experimental}

Single crystals of \tmc\ grown using an electro-crystallization technique~\cite{Bechgaard1981JAmChemSoc}
were provided from K.~Bechgaard.
We have measured $\Rzz$ of about 10 samples.
We report here the results for Sample \#1 (approximately $2.0\times 0.2\times 0.1$~mm$^3$) % Sample 0611-#1
and Sample \#2 (approximately $1.4\times 0.3\times 0.1$~mm$^3$). % Sample 0605-#4. 
We note that Sample \#1 is identical to the sample used in the previous report~\cite{Yonezawa2008.PhysRevLett.100.117002}.
We measured $\Rzz$ using a conventional AC four-probe method with a lock-in amplifier
with frequencies 277~Hz or 887~Hz.
In order to attach electrical probes, we first evaporated gold pads on the $ab$ surfaces of the \tmc\ crystal,
and then glued gold wires with silver paste to the gold pads. 
The measurements were performed with a {}$^4$He-{}$^3$He dilution refrigerator,
which allows measurements down to 80~mK.
Temperature was measured with a RuO$_2$ resistance thermometer,
for which the magnetoresistance had been calibrated.
The sample was first slowly cooled ($\sim 0.1$--$0.2$~K/min) from room temperature to 77~K 
in order to prevent the sample from cracking caused by the large thermal expansion of molecular crystals.
The anion ordering temperature of \tmc\ is 24~K, 
at which tetrahedra of ClO\sub{4} align up and down alternatively to the $b\dash$ direction.
Therefore, in the temperature interval between 25~K and 22~K,
a cooling rate as slow as 2--4~mK/min was chosen 
to ensure that anions are well ordered and the whole sample is in the ``relaxed state''.

Magnetic fields are applied using the ``Vector Magnet'' system~\cite{Deguchi2004RSI}.
The directions of the orthogonal crystalline axes (the $a$, $b\dash$, and $\cstar$ axes) 
of the sample were determined from the anisotropy of $\Hcc$ at 0.1~K.
The accuracy of the field alignment with respect to the $ab\dash$ plane 
and of the $a$ axis within the $ab\dash$ plane are both better than 0.1~degree.
We also determined the directions of the triclinic crystalline axes (the $b$ and $c$ axes) for Sample \#1
from angular magnetoresistance oscillations.
The details of these procedures are presented elsewhere~\cite{Kusaba2007SolidStateSci}.
For Sample \#2, the directions of the triclinic axes are estimated
by comparing its magnetoresistance with that of Sample \#1.
Below, we denote 
the polar angle between {\boldmath$H$} and the $\cstar$ axis as $\theta$, and 
the azimuthal angle within the $ab\dash$ plane as $\phi$
which is measured from the $a$ axis.
We define $\phi$ so that the $b$ axis lies in the quadrant $0\deg<\phi<90\deg$
as indicated with the arrows at the top of Fig.~\ref{fig:Tc-angle}.

%%%%%%%%%%%%%%%%%%%%%%%%%%%%%%%%%%%%%%% Results %%%%%%%%%%%%%%%%%%%%%%%%%%%%%%%%%%%%%%%%%%%

\section{Results}\label{sec:Results}

\subsection{Sample Characterization}

%%%%%%%%%%%%%%%%%%%%%%%%%%%%%%%%%%%%%%%%%%%%%%%%
\begin{figure}
\begin{center}
\includegraphics[width=7.5cm]{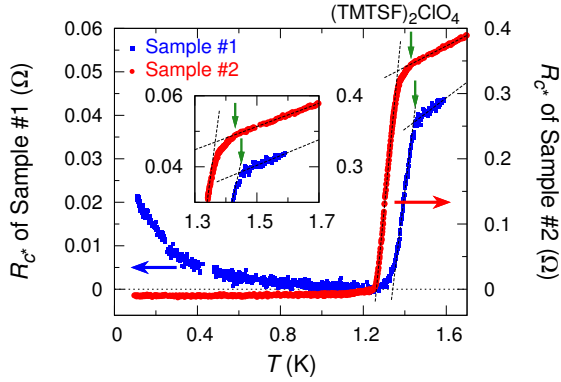}
\caption{(color online)
Temperature dependence of $\Rzz$ in zero field for Sample \#1 (blue squares) and Sample \#2 (red circles).
The onset temperatures are 1.45~K for Sample \#1 and 1.43~K for Sample \#2,
indicated with the small green arrows. 
The inset shows $\Rzz(T)$ near the SC onset.
The broken lines are linear extrapolations of the normal state resistance
and the resistance in the superconducting transitions.
The intersections of these lines correspond to $\Tcj$.
\label{fig:zero-resistance}}
\end{center}
\end{figure}
%%%%%%%%%%%%%%%%%%%%%%%%%%%%%%%%%%%%%%%%%%%%%%%%

We first present $\Rzz(T)$ in zero field in Fig.~\ref{fig:zero-resistance}.
Sample \#1 exhibits an onset of superconductivity at as high as 1.45~K and zero resistance at 1.30~K,
while Sample \#2 started to show a drop in $\Rzz(T)$ at 1.43~K and zero resistance at 1.26~K.
As one can see, $\Rzz$ of Sample \#1 increases 
again in the superconducting state.
This increase, which was almost independent of magnetic fields, 
is attributable to small cracks in the sample.
As we will explain later, this extrinsic increase of resistance does not affect our analysis.

We can quantitatively compare the transition temperatures in zero field, $\Tcz$, with the sample quality 
using the formula~\cite{Radtke1993PhysRevB}
\begin{align}
\ln\left(\frac{\Tcz\spsm{clean}}{\Tcz}\right) = 
\Psi\left( \frac{1}{2} + \frac{\alpha\Tcz\spsm{clean}}{2\pi\Tcz} \right)
- \Psi\left( \frac{1}{2} \right),
\label{eq:digamma}
\end{align}
where $\Psi(x)$ is the digamma function, 
$\Tcz\spsm{clean}$ is the ideal $\Tcz$ in the limit of no impurities, 
$\alpha=\hbar/2\tau k\subm{B} \Tcz\spsm{clean}$ is the depairing parameter for the isotropic impurity scattering, 
and $\tau$ is the scattering relaxation time.
In the study of the impurity-concentration dependence of $\Tcz$ by Joo \etal~\cite{Joo2005},
they obtained $\Tcz\spsm{clean}=1.57$~K using $\Tcz$ determined from the interception
of the linear extrapolations of the normal state resistance and the resistance in the SC transition,
which we denote as $\Tcjz$ following the notation by Lee \etal~\cite{Lee1997}. 
Using $\Tcjz=1.44$~K for Sample \#1 and 1.37~K for Sample \#2
as shown in the inset of Fig.~\ref{fig:zero-resistance},
we obtain $\tau_1=2.3\times 10^{-11}$~sec, $\tau_2=1.5\times 10^{-11}$~sec,
and the ratio $\tau_1/\tau_2=1.5$, 
where $\tau_1$ and $\tau_2$ are $\tau$ for Sample \#1 and Sample \#2, respectively.

%%%%%%%%%%%%%%%%%%%%%%%%%%%%%%%%%%%%%%%%%%%%%%%%
\begin{figure}
\begin{center}
\includegraphics[width=7.5cm]{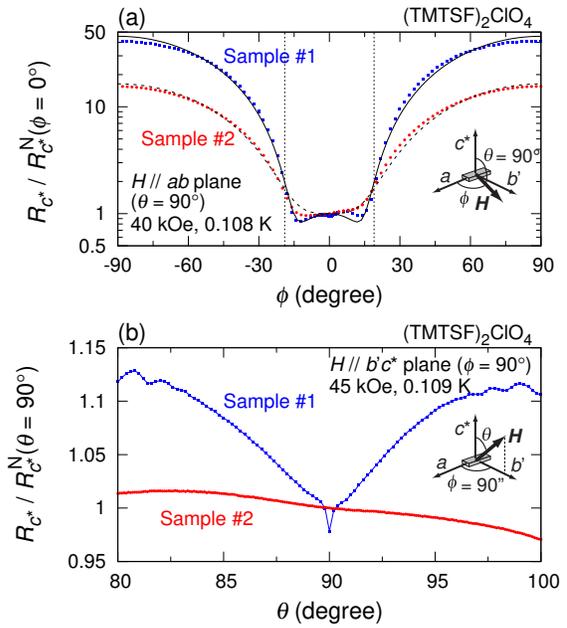}
\caption{(color online)
AMRO of Sample \#1 (blue squares) and Sample \#2 (red circles).
(a) AMRO for magnetic fields parallel to the $ab$ plane.
The vertical axis is in a logarithmic scale.
Resistance is normalized by the resistance without the contribution of superconductivity 
$\Rzz^{\mathrm N}$ at $\phi=0\deg$ (\Hvec${}\parallel a$).
The solid and broken curves are results of fitting of Eqs.~\eqref{eq:AMRO1}--\eqref{eq:AMRO3} to the data.
The vertical dotted lines represent $\pm\phidco=\pm (19\pm 1)\deg$, which we defined as the angle above which
$\Rzz(T)$ in the normal state above 20~kOe exhibits non-metallic behavior for Sample \#1.
%The dips of the curve for Sample \#1 at $\phi\sim \pm\phi\subm{TAE}=\pm 13\deg$ is due to the third angle effect.
(b) AMRO for magnetic fields parallel to the $b\dash\cstar$ plane.
Resistance is normalized by $\Rzz^{\mathrm N}$ at $\theta=90\deg$ (\Hvec${}\parallel b\dash$).
The dip of $\Rzz(\theta)$ at $\theta=90\deg$ for Sample \#1 is a contribution of superconductivity.
\label{fig:AMRO}}
\end{center}
\end{figure}
%%%%%%%%%%%%%%%%%%%%%%%%%%%%%%%%%%%%%%%%%%%%%%%%

It is important to note that 
the difference in sample quality strongly affects 
the amplitude of the angular magnetoresistance oscillations (AMRO) 
in the normal state plotted in Fig.~\ref{fig:AMRO}.
The AMRO can be understood within the framework of 
a semi-classical motion of carriers on the Fermi surface~\cite{Kang1992,Danner1994,Osada1996}.
For an occurrence of AMRO, $\tau$ should be long enough
so that carriers can ``feel'' the shape of the Fermi surface before being scattered.
Therefore, $\omega\subm{c}\tau$ larger than 1 is necessary for a large amplitude of AMRO,
where $\omega\subm{c}=ev\subm{F}\mu_0H c/\hbar$ is the cyclotron frequency of the carriers,
$e$ the elemental charge, $v\subm{F}$ the Fermi velocity along the $a$ direction, and $c$ the lattice constant.
Sugawara \etal~\cite{Sugawara2006} recently obtained an analytical expression of $\Rzz(H,\phi)$
for an $ab$-plane field rotation within an orthorhombic tight-binding model:
\begin{align}
\frac{\Rzz(H, \phi)}{\Rzz(H=0)}%
&= \frac{ \sqrt{2} g(\omega\subm{c}\tau,\,\phi) } %
{\sqrt{ f(\omega\subm{c}\tau,\,\phi) +  g(\omega\subm{c}\tau,\,\phi) }}\ \ ,\label{eq:AMRO1}\\
f(\omega\subm{c}\tau,\,\phi)&\equiv 1-(\omega\subm{c}\tau)^2(\sin^2\phi-\eta^2\cos^2\phi)\ ,\label{eq:AMRO2}\\
g(\omega\subm{c}\tau,\,\phi)&\equiv \sqrt{ f(\omega\subm{c}\tau,\,\phi)^2 + (2\omega\subm{c}\tau\sin\phi)^2}\ \label{eq:AMRO3},
\end{align}
where $\eta$ is the anisotropy parameter of the velocity between the $a$ and the $b\dash$ directions
and was estimated to be $\eta= 0.221$--$0.265$ for their samples.
In these equations, $\phi$ dependence of $\Rzz$ disappears as $\omega\subm{c}\tau\to 0$,
explaining indeed the importance of a large $\omega\subm{c}\tau$ for an occurrence of the AMRO.

As shown in Fig.~\ref{fig:AMRO}, 
$\Rzz$ of Sample \#1 exhibits an oscillation with a large amplitude
for both field rotation in the $ab$ plane and rotation in the $b\dash\cstar$ plane,
while the oscillation of $\Rzz$ of Sample \#2 is rather smaller.
The difference of $\tau$ can be estimated by fitting Eqs.~\eqref{eq:AMRO1}--\eqref{eq:AMRO3} to the data,
by using $\omega\subm{c}\tau$ and $\eta$ as fitting parameters.
The results of the fitting are shown with the solid and broken curves in Fig.~\ref{fig:AMRO}(a).
From this fitting we obtained $\omega\subm{c}\tau_1 = 13$ and $\eta=0.27$ for Sample \#1,
and $\omega\subm{c}\tau_2 = 5.0$ and $\eta=0.24$ for Sample \#2 at 40~kOe.
Using the value $v\subm{F}=1.8\times 10^{5}$~m/sec 
obtained from a microwave conductance experiment~\cite{Kovalev2003JApplPhys},
we estimated $\omega\subm{c} = 1400$~GHz at 40~kOe.
From these values, $\tau_1$ can be estimated to be $9.0\times 10^{-12}$~sec,
$\tau_2$ to be $3.5\times 10^{-12}$~sec,
and the ratio $\tau_1/\tau_2$ to be 2.6.
In addition, we evaluated the mean free path $l=v\subm{F}\tau$ to be $1.6\times 10^{-6}$~m for Sample \#1
and $6.1\times 10^{-7}$~m for Sample \#2,
demonstrating that these samples are extremely clean.
The results of these analyses are summarized in Table~\ref{tab:samples}.
We note that there is a SC contribution in the $\Rzz(\phi)$ data for Sample \#1 even at 45~kOe
and the amplitude of the $\Rzz(\phi)$ oscillation without a SC contribution 
is probably larger than Fig.~\ref{fig:AMRO}.
Thus the intrinsic $\omega\tau_1$ should be a bit larger than the present analysis.

In the above analyses, we have assumed the isotropic ($\kbm$-independent) impurity scattering.
An anisotropic component of the impurity scattering contributes 
differently to the suppression of $\Tc$ and to the AMRO.
An inclusion of the anisotropic impurity scattering 
may lead to \textit{weaker} suppression of $\Tc$ of a $d$-wave superconductor
than the case of the isotropic impurity scattering~\cite{Haran1996.PhysRevB.54.15463,Lin1999.PhysRevB.59.6047}.
Meanwhile, if the scattering time is $\kbm$-dependent,
$\tau$ in Eqs.~\eqref{eq:AMRO1}--\eqref{eq:AMRO3} should be replaced by the averaged value of $\tau(\kbm)$ along 
the path of the cyclotron motion.
Thus the anisotropic impurity scattering
should suppress the amplitude of the AMRO as the isotropic scattering does.
To summarize the above discussion, the anisotropy in impurity scattering may weaken the suppression of $\Tc$
and result in an effectively longer $\tau$ in Eq.~\eqref{eq:digamma},
while the $k$-averaged shorter $\tau$ should appear fully in Eqs.~~\eqref{eq:AMRO1}--\eqref{eq:AMRO3}.
The fact that $\tau$ obtained from Eq.~\eqref{eq:digamma} is longer than $\tau$ from the analysis of the AMRO
may be attributed to the existence of anisotropy in the impurity scattering in our samples.

It should be noticed that the AMRO in the $ab$ plane in Fig.~\ref{fig:AMRO}(a) is closely related to 
the suppression of the inter-layer hopping due to the FIDC.
The small magnetoresistance ratio $\Rzz(\Hbm\parallel b\dash)/\Rzz^{\mathrm N}(\Hbm\parallel a)$ 
for Sample \#2 indicates that
the carriers in Sample \#2 are not well confined in the $ab$ plane; i.\,e.
the dimensionality of the electronic system for Sample \#2 is not well reduced in the present field range.

%%%%%%%%%%%%%%%%%%%%%%%%%%%%%%%%%%%%%%%%%%%%%%%%
\begin{table}
\begin{center}
\caption{Summary of the quality and superconducting parameters of the samples of \tmc.
\label{tab:samples}}
\begin{tabular}{ccc}\hline
 & Sample \#1 & Sample \#2 \\ \hline
$\Tco$ at $H=0$   & 1.45~K & 1.43~K \\
$\Tcj$ at $H=0$   & 1.44~K & 1.37~K \\
$\Tc^{R=0}$ at $H=0$  & 1.30~K & 1.26~K \\
$\tau$ obtained from Eq.\eqref{eq:digamma} & $2.3\times 10^{-11}$~sec & $1.5\times 10^{-11}$~sec \\
$\tau$ obtained from Eqs.\eqref{eq:AMRO1}-\eqref{eq:AMRO3} & $9.0\times 10^{-12}$~sec & $3.5\times 10^{-12}$~sec \\
$l=v\subm{F}\tau$ obtained from Eqs.\eqref{eq:AMRO1}--\eqref{eq:AMRO3} & $1.6\times 10^{-6}$~m & $6.1 \times 10^{-7}$~m \\
%$\Rzz(\Hbm\parallel b\dash)/\Rzz\spsm{N}(\Hbm\parallel a)$ at 40~kOe, 0.1~K&  41 & 15  \\ 
\hline
$\Horb$ for $\Hbm\parallel a$   &   70~kOe  &  60~kOe \\
$\Horb$ for $\Hbm\parallel b\dash$   &   38~kOe  &  28~kOe \\
$\Horb$ for $\Hbm\parallel \cstar$   &   1.5~kOe  &  1.5~kOe \\
$\xi_a(0)$ & 450~\AA & 490~\AA \\
$\xi_{b\dash}(0)$ & 240~\AA & 230~\AA \\
$\xi_{\cstar}(0)$ & 10~\AA & 13~\AA \\ \hline
$\ta$ & 1200~K & 1200~K  \\
$\tb$ & 310~K  & 290~K   \\
$\tc$ & 7.0~K  & 8.5~K   \\ \hline
\end{tabular}
\end{center}
\end{table}
%%%%%%%%%%%%%%%%%%%%%%%%%%%%%%%%%%%%%%%%%%%%%%%%

%Hc2_orb; Sample #1
%H//a: 70.3107901405002 kOe
%H//b': 37.8292846987125 kOe
%H//c*: 1.5830738316639 kOe
%Hc2_orb; Sample #2
%H//a: 59.5794122286558 kOe
%H//b': 28.1340892344725 kOe
%H//c*: 1.52154600559803 kOe    

%%%%%%%%%%%%%%%%%%%%%%%%%%%%%%%%%%%%%%%%%%%%%%%%
\begin{figure}
\begin{center}
\includegraphics[width=7.5cm]{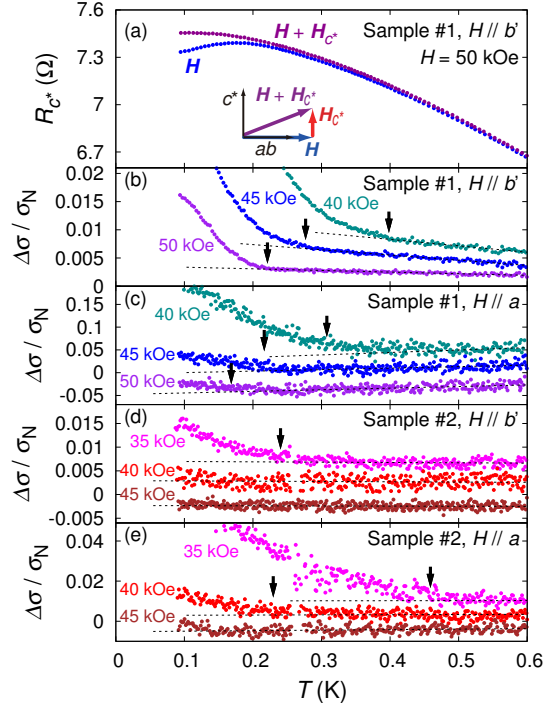}
\caption{(color online)
Examples of the definition of $\Tco$.
(a) $\Rzz(T)$ of Sample \#1 for \Hvec${}\parallel b\dash$ at 50~kOe.
The lower curve is the result for $H_{\cstar}=0$~kOe and the upper curve is for $H_{\cstar}=0.5$~kOe.
The large difference of these curves at low temperatures is attributable to
a contribution of superconductivity.
(b)--(e) 
Conductance difference $\Delta\sigma(T; \Hbm)\equiv \Rzz^{-1}(T, \Hbm) - \Rzz^{-1}(T, \Hbm+\Hbm_{\cstar})$
normalized by the normal-state conductance $\sigma\subm{N}$ at $\Tco$.
The small arrows indicate $\Tco$, where $\Delta\sigma(T)$ starts to increase from 
the linear temperature dependence at higher temperatures.
The dotted lines are the results of fitting with a linear function of temperature to 
$\Delta\sigma(T)/\sigma\subm{N}$ at high temperatures, which indicate the normal-state behavior.
We note that some curves are shifted vertically for clarity.
(b) $\Delta\sigma(T)/\sigma\subm{N}$ of Sample \#1 for \Hvec${}\parallel b\dash$ for
$(H_{b\dash}, H_{\cstar}) = $ (40~kOe, 1.0~kOe) (dark cyan), 
(45~kOe, 1.0~kOe) (blue), and
(50~kOe, 0.5~kOe) (purple).
(c) $\Delta\sigma(T)/\sigma\subm{N}$ of Sample \#1 for \Hvec${}\parallel a$ for
$(H_{a}, H_{\cstar}) = $ (40~kOe, 1.0~kOe) (dark cyan), 
(45~kOe, 1.0~kOe) (blue), and
(50~kOe, 1.0~kOe) (purple).
(d) $\Delta\sigma(T)/\sigma\subm{N}$ of Sample \#2 for \Hvec${}\parallel b\dash$ for
$(H_{b\dash}, H_{\cstar}) = $ (35~kOe, 0.5~kOe) (magenta), 
(40~kOe, 0.5~kOe) (red), and
(45~kOe, 1.0~kOe) (brown).
(e) $\Delta\sigma(T)/\sigma\subm{N}$ of Sample \#2 for \Hvec${}\parallel a$ for
$(H_{a}, H_{\cstar}) = $ (35~kOe, 0.5~kOe) (magenta), 
(40~kOe, 0.5~kOe) (red), and
(45~kOe, 0.5~kOe) (brown).
\label{fig:example}}
\end{center}
\end{figure}
%%%%%%%%%%%%%%%%%%%%%%%%%%%%%%%%%%%%%%%%%%%%%%%%

\subsection{Definition of $\Tco$}

Temperature dependence of $\Rzz$ of Sample \#1 in a 50-kOe magnetic field parallel to the $b\dash$ axis
is plotted with the lower curve in Fig.~\ref{fig:example}(a).
We observed decrease of $\Rzz(T)$ when cooled below 0.2~K, 
which is consistent with previous reports~\cite{Lee1995, Naughton1997}.
In order to confirm that this decrease is attributable to a contribution of superconductivity,
we measured $\Rzz(T)$ after adding a small out-of-plane magnetic field $\Hbm_{\cstar}=(0,0,H_{\cstar})$
where $H_{\cstar}=0.5$--$1.0$~kOe.
If the decrease of $\Rzz(T)$ is due to a superconductivity, $\Hbm_{\cstar}$ should suppress the superconductivity and 
eliminate the decrease of $\Rzz(T)$. 
As plotted in Fig.~\ref{fig:example}(a), the addition of $\Hbm_{\cstar}$ indeed eliminated 
the decrease of $\Rzz(T)$.
Therefore, we confirm that the decrease of $\Rzz(T)$ is a precursor of the superconductivity.
We defined $\Tco$ based on the same idea.
We evaluated the conductance difference 
$\Delta\sigma(T; \Hbm)\equiv \Rzz^{-1}(T; \Hbm)-\Rzz^{-1}(T; \Hbm+\Hbm_{\cstar})$
and defined $\Tco$ as the temperature where $\Delta\sigma(T)$ exhibits a sharp increase,
as shown with the small arrows in Fig.~\ref{fig:example}(b)--(e).
This definition characterizes the very onset of the superconductivity,
or probably the onset of a vortex liquid state.
In the early reports of \tmc\ for $\Hbm\parallel b\dash$~\cite{Lee1995, Naughton1997}
$\Tc$ was defined as the temperature of the peak of $\Rzz(T)$.
However, their definition is not appropriate for our study,
because the peak definition cannot be applied when field is near the $a$ direction,
where a peak of $\Rzz(T)$ is absent.
The $\Tcj$ definition~\cite{Lee1997} cannot be easily applied either, 
because of the complicated non-linear temperature dependence of $\Rzz$ in the normal state in high fields.
Meanwhile, our definition keeps consistency in any in-plane field angle $\phi$, thus fits well to our study.
This definition also has an advantage that 
$\Tco$ is not affected by the extrinsic additional resistance
due to small cracks in the sample, since it is cancelled in the subtraction.
Note that all the $\Tco$ data presented in this paper are obtained from temperature increasing sweeps.

We also studied $\Tco$ for $\theta\ne 90\deg$, \textit{i.\,e.}
for fields titled from the $ab$ plane and for fields parallel to the $\cstar$ axis.
The definition of $\Tco$ for $\theta\ne 90\deg$ is 
a straightforward extension of the definition for $\theta=90\deg$:
$\Tco(H, \theta, \phi)$ is the temperature where $\Delta\sigma(T; \Hbm)\equiv \Rzz^{-1}(T; \Hbm)-\Rzz^{-1}(T; \Hbm+\Hbm_{\cstar})$
exhibits a slope change, where $\Hbm=(H\sin\theta\cos\phi, H\sin\theta\sin\phi, H\cos\theta)$
and $\Hbm_{\cstar}=(0,0,H_{\cstar})$ ($H_{\cstar}=0.5$--$3.0$~kOe).

We note that the anomalies in $\Delta\sigma(T)$ in Fig.~\ref{fig:example}(b)--(e)
are not due to the normal state magnetoresistance,
because it is unlikely that an abrupt change in the difference between 
$\Rzz(T; \Hbm)$ and $\Rzz(T; \Hbm+\Hbm_{\cstar})$ occurs at a certain temperature 
only from the magnetoresistance.
In addition, since $H_{\cstar} \ll H$ 
the tilt angle between $\Hbm$ and $\Hbm+\Hbm_{\cstar}$ is less than $1\deg$ in high fields,
which is smaller than the typical angle scale of the AMROs. 

It is clear from Fig.~\ref{fig:example}(b) that
$\Tco$ of Sample \#1 remains finite even at 50~kOe for $\Hbm\parallel b\dash$.
Interestingly, for $\Hbm\parallel a$ 
a small upturn of $\Delta\sigma(T)$ remains visible at 50~kOe for Sample \#1,
which indicate that $\Tco$ is finite also at 50~kOe for $\Hbm\parallel a$.
On the contrary, for Sample \#2 an anomaly of $\Delta\sigma(T)$ was absent
both for $\Hbm\parallel a$ and $\Hbm\parallel b\dash$ above 45~kOe.
These results suggest large differences between the $H$-$T$ phase diagrams of the two samples
both for $\Hbm\parallel a$ and for $\Hbm\parallel b\dash$,
especially in low-temperature high-field regions.

\subsection{$H$-$T$ and $T$-$\phi$ Phase diagrams for in-plane fields}\label{subsec:phase_diagram}

%%%%%%%%%%%%%%%%%%%%%%%%%%%%%%%%%%%%%%%%%%%%%%%%
\begin{figure}
\begin{center}
\includegraphics[width=7.5cm]{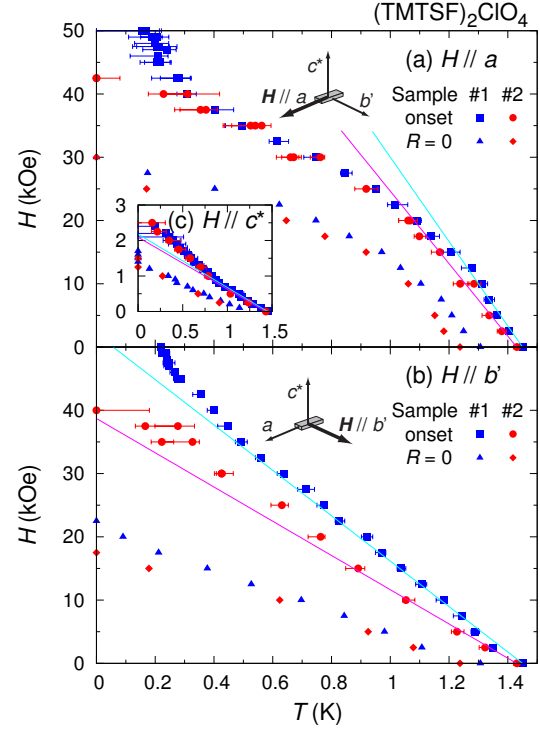}
\caption{(color online)
Magnetic fields v.s. $\Tco$ of Sample \#1 (blue squares) and Sample \#2 (red circles)
(a) for $\Hbm\parallel a$, 
(b) for $\Hbm\parallel b\dash$, and
(c) for $\Hbm\parallel \cstar$.
The solid lines are the results of linear fitting near $H=0$.
The temperatures where $\Rzz(T)-\Rzz^{\mathrm{crack}}(T)$ becomes zero, $\Tc^{R=0}$,
are also plotted for Sample \#1 (triangles) and for Sample \#2 (diamonds).
\label{fig:phase_diagram}}
\end{center}
\end{figure}
%%%%%%%%%%%%%%%%%%%%%%%%%%%%%%%%%%%%%%%%%%%%%%%%

The resulting $H$-$T$ phase diagrams of both Sample \#1 and \#2 
for $\Hbm\parallel a$, $\Hbm\parallel b\dash$, and $\Hbm\parallel \cstar$
are presented in Fig.\ref{fig:phase_diagram}.
In this figure, we also plotted $\Tc^{R=0}$, at which $\Rzz(T)-\Rzz^{\mathrm{crack}}(T)$ becomes zero,
where $\Rzz^{\mathrm{crack}}(T)$ is the resistance due to small cracks in the samples.
The region between the $\Tco$ curve and the $\Tc^{R=0}$ curve 
probably corresponds to a vortex liquid state.

For $\Hbm\parallel a$ and $\Hbm\parallel \cstar$, 
the onset curves of Sample \#1 and Sample \#2 nearly coincide with each other,
except for the low-temperature high-field region for $\Hbm\parallel a$.
However, for $\Hbm\parallel b\dash$ the two curves are quite different
and this difference starts to develop at $H=0$.
These different sample dependences of the shape of the curves are discussed in the next section.
We note that above 0.5~K these curves agree qualitatively with
those reported by Murata \etal~\cite{Murata1987},
and the curve of Sample \#1 for $\Hbm\parallel b\dash$ is qualitatively similar to
the curve reported by Oh and Naughton~\cite{Oh2004},
although we used a different definition of the onset temperature.

%%%%%%%%%%%%%%%%%%%%%%%%%%%%%%%%%%%%%%%%%%%%%%%%
\begin{figure}
\begin{center}
\includegraphics[width=7.5cm]{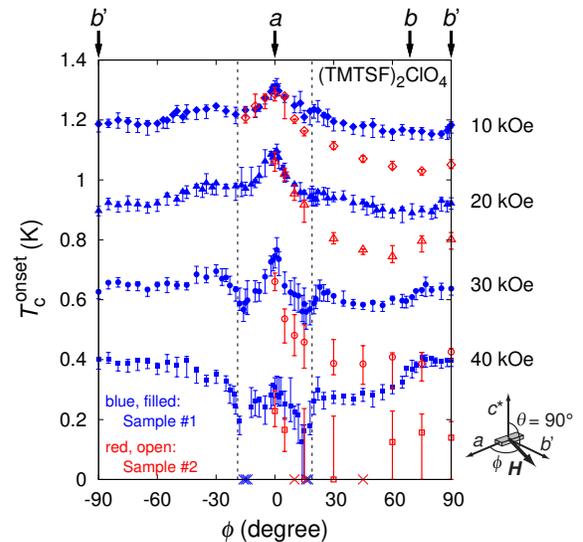}
\caption{(color online)
In-plane field-angle dependence of $\Tco$ in various magnitude of magnetic field
for Sample \#1 (filled blue symbols) and for Sample \#2 (open red symbols)
at 10~kOe (diamonds), at 20~kOe (triangles), at 30~kOe (circles) and at 40~kOe (squares).
%We note that we have not checked the direction of the $b$ axis of Sample \#2.
%Therefore it is not clear whether $\phi>0$ for Sample \#2 corresponds $\phi>0$ for Sample \#1 or not.
The directions of some crystalline axes are illustrated by the arrows at the top of the graph.
The vertical broken lines represent the angle where the FIDC occurs, $\pm\phidco = \pm 19\deg$.
Inside these lines the electronic state is anisotropic 3D 
whereas the dimensionality starts to be lowered in the outside.
\label{fig:Tc-angle}}
\end{center}
\end{figure}
%%%%%%%%%%%%%%%%%%%%%%%%%%%%%%%%%%%%%%%%%%%%%%%%

We then plotted in Fig.~\ref{fig:Tc-angle} the in-plane field-angle $\phi$ dependence of
$\Tco$ under magnetic fields of various strengths,
for both Sample \#1 and \#2.
The discrepancy between the curves of Sample \#1 and Sample \#2
is small near $\Hbm\parallel a$, whereas the discrepancy starts to be larger as $|\phi|$ increases beyond $20\deg$.
The curves at 10~kOe, with a sharp peak at $\phi=0\deg$ 
and a broad minimum around $\phi=\pm 90\deg$,
are qualitatively consistent with the $\Hcc(\phi)$ curve at 1.03~K obtained by Murata \etal~\cite{Murata1987}.

However, in higher fields, the curves for Sample \#1 exhibit various anomalous structures.
One of the anomalies is the dips of $\Tco(\phi)$ around $\phi=\phidip=\pm 17\deg$
observed above 20~kOe.
We argue that these dips originate from the FIDC, as will be discussed in Sec.~\ref{subsec:FIDC}.
 
\begin{figure}
\begin{center}
\includegraphics[width=7.5cm]{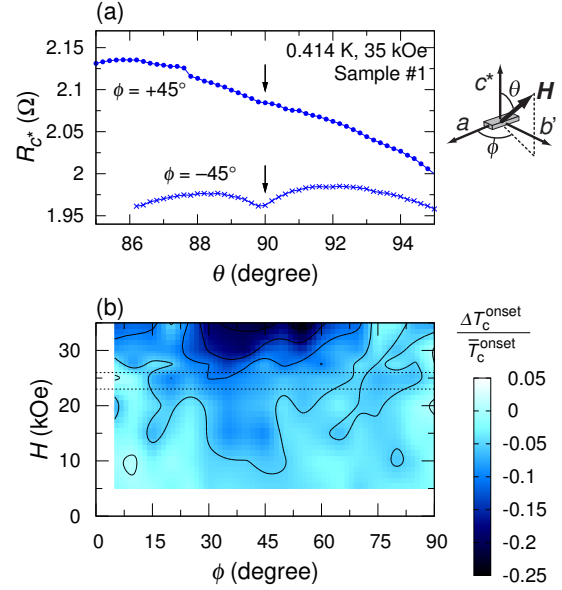}
\caption{(color online)
(a) Polar angle $\theta$ dependence of $\Rzz$ for Sample \#1 at $\phi=+45\deg$ (circles)
and at $\phi=-45\deg$ (crosses)
at $T=0.4$~K and $H=35$~kOe.
In this condition, the temperature is well below $\Tco$ for $\phi=-45\deg$ 
while it is approximately equal to $\Tco$ for $\phi=+45\deg$.
The arrows indicate the $ab$ plane, where dip of $\Rzz(\theta)$ due to superconductivity 
should be observed if $\Tco(\phi)>0.4$~K.
A dip was observed for $\phi=-45\deg$ but absent for $\phi=+45\deg$,
proving that the asymmetry of $\Tco(\phi)$ is not because of the misalignment of the field to the $ab$ plane.
We note that the different behavior of the normal state magnetoresistance 
originates from the triclinic crystal structure:
In the present case, the fact that the angle between the $b$ axis and $\Hbm$ for $\phi=+45\deg$
is different from the angle between the $b$ axis and $\Hbm$ for $\phi=-45\deg$ 
leads to the different behavior of $\Rzz(\theta)$.
(b) Contour plot of $\Delta\Tco(\phi)/\bar{T}\subm{c}\spsm{onset}(\phi)$, 
where $\Delta\Tco(\phi) \equiv \Tco(\phi) - \Tco(-\phi)$ and 
$\bar{T}\subm{c}\spsm{onset}(\phi) \equiv (\Tco(\phi) + \Tco(-\phi))/2$.
The contours are drawn with an interval of 0.05.
This quantity represents a breaking of the mirror symmetry of the $\Tco(\phi)$ curve 
with respect to the $a$ axis and to the $b\dash$ axis.
The dotted lines represent the range of the estimated value of the Pauli limit $\Hp\sim23$--26~kOe.
\label{fig:asymmetry}}
\end{center}
\end{figure}

Another anomaly of $\Tco(\phi)$ is the shoulder-like structure at $\phi\sim+70\deg$. 
This deserves a special attention, because it is observed only for $\phi>0\deg$ but \textit{not} for $\phi<0\deg$,
resulting in a breaking of the mirror symmetry of $\Tco(\phi)$ with respect to the $a$ and the $b\dash$ axes.
This asymmetry is clearly not due to the misalignment of the magnetic field to the crystalline axes
nor the existence of several crystal domains with slightly tilted crystalline axes,
because the $\theta$ dependence of $\Rzz$ plotted in Fig.~\ref{fig:asymmetry}(a) 
exhibits a clear dip at \Hvec${}\parallel ab$ for $\phi=-45\deg$ 
while a dip is absent for $\phi=+45\deg$.
If the asymmetry were due to the misalignment or due to the existence of crystal domains,
dips of $\Rzz(\theta)$ with the same depth
would have been observed both for $\phi=+45\deg$ and $\phi=-45\deg$
and at least one of them should not be centered at $\theta=90\deg$.
In order to verify at which field this asymmetry appears,
we plotted $\Delta\Tco(\phi)/\bar{T}\subm{c}\spsm{onset}(\phi)$ 
as a contour plot against $\phi$ and $H$ in Fig.~\ref{fig:asymmetry}(b), 
where $\Delta\Tco(\phi)\equiv \Tco(\phi) - \Tco(-\phi)$ and 
$\bar{T}\subm{c}\spsm{onset}(\phi)$ is the average of $\Tco(\phi)$ and $\Tco(-\phi)$.
A finite value of $\Delta\Tco/\bar{T}\subm{c}\spsm{onset}$
represents the existence of asymmetry of the $\Tco(\phi)$ curve with respect to $\phi=0\deg$ and $90\deg$.
It is clear that $|\Delta\Tco|/\bar{T}\subm{c}\spsm{onset}$ suddenly increases 
above 25--30~kOe as field increases.
This fact indicates that the asymmetry is nearly absent at lower fields but starts to emerge at 25--30~kOe.
Therefore, the asymmetry cannot be attributed to conventional origins such as 
anisotropy of the Fermi velocity or the  effective mass on the Fermi surface,
because asymmetries from such origins should develop from $H=0$.
Possible origins of the asymmetry will be discussed in Sec.~\ref{subsec:Pauli_Hb}.

%%%%%%%%%%%%%%%%%%%%%%%%%%%%%%%%%%%%%%%%%%%%%%%%
%%%%%%%%%%%%%%%%%%%%%%%%%%%%%%%%%%%%%%%%%%%%%%%%

\subsection{$\Tco$ for fields titled from the $ab$ plane}\label{subsec:phase_diagram_tilt}

\begin{figure}
\begin{center}
\includegraphics[width=7.5cm]{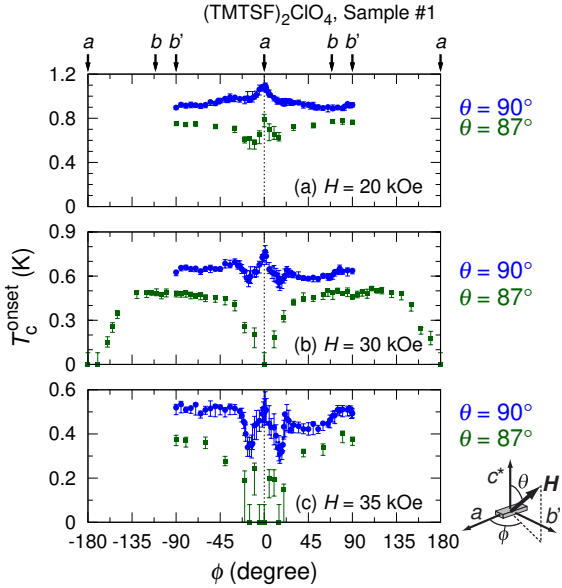}
\caption{(color online)
Field-angle $\phi$ dependence of $\Tco$ for Sample \#1 for $\theta=90\deg$ ($\Hbm\parallel ab$) (blue circles)
and $\theta =87\deg$ (green squares) at (a) 20~kOe, (b) 30~kOe, and (c) 35~kOe.
\label{fig:Tc-angle_87deg}}
\end{center}
\end{figure}

\begin{figure}
\begin{center}
\includegraphics[width=7.5cm]{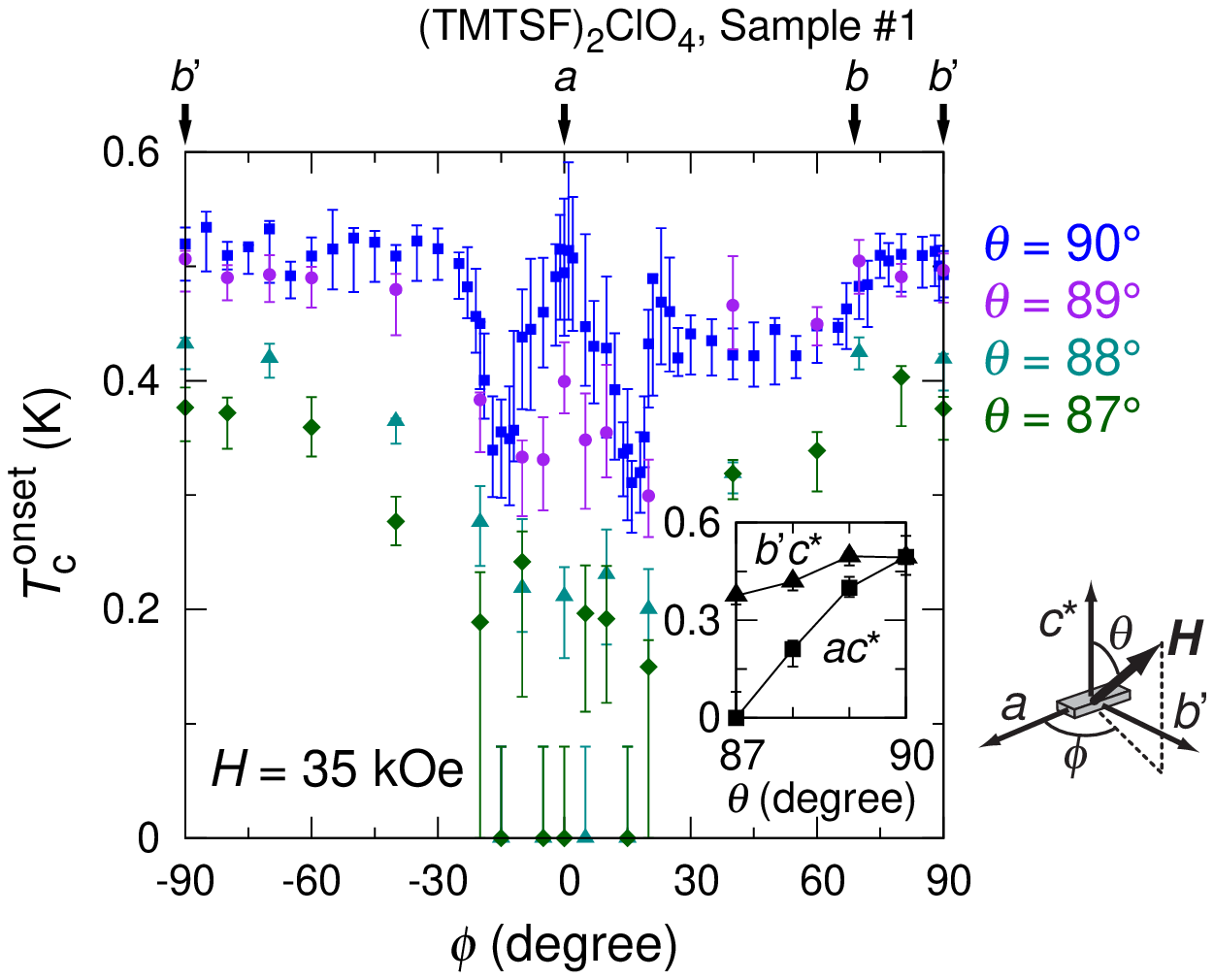}
\caption{(color online)
Field-angle $\phi$ dependence of $\Tco$ for Sample \#1 at $H=35$~kOe 
for $\theta=90\deg$ (blue squares) ($\Hbm\parallel ab$),
$89\deg$ (purple circles), $88\deg$ (dark-cyan triangles), and $87\deg$ (green diamonds).
The inset is the $\theta$ dependence of $\Tco$ at $\phi=0\deg$ ($\Hbm\parallel a\cstar$; squares) 
and at $\phi=90\deg$ ($\Hbm\parallel b\dash\cstar$; triangles).
\label{fig:Tc-angle-theta_35kOe}}
\end{center}
\end{figure}

We also studied $\Tco$ for Sample \#1 in magnetic fields 
slightly tilted from the conductive $ab$ plane, \textit{i.\,e.} $\theta\ne 90\deg$.
We present in Fig.~\ref{fig:Tc-angle_87deg} a comparison of the data for $\theta=87\deg$
with those for $\theta=90\deg$ ($\Hbm\parallel ab$).
We also show in Fig.~\ref{fig:Tc-angle-theta_35kOe} the $\theta$ dependence of $\Tco(\phi)$ at 35~kOe.
Strictly speaking, a two-fold rotational symmetry can be absent in $\Tco(\phi)$ for $\theta\ne 90\deg$.
Hence we checked $\Tco(\phi)$ in the range $-180\deg\le \phi \le 180\deg$ at $H=30$~kOe and $\theta=87\deg$
as shown in Fig.~\ref{fig:Tc-angle_87deg}(b),
which however does not indicate clear breaking of a two-fold symmetry in $\Tco(\phi)$
within the present experimental accuracy.

It is clear from these figures that in high fields 
the reduction of $\Tco$ due to the tilt of the magnetic field
depends largely on $\phi$.
At 20~kOe, the reduction of $\Tco$ for $\Hbm\parallel a$ and for $\Hbm\parallel b\dash$ 
does not differ very much.
However, above 30~kOe, $\Tco$ at $\Hbm\parallel a$ is completely reduced to zero 
when the field is tilted $3\deg$ from the $ab$ plane,
whereas $\Tco$ for $\Hbm\sim b\dash$ is still finite at $\theta=87\deg$.
The difference between the reduction of $\Tco(\theta)$ for $\Hbm\parallel a$ and for $\Hbm\parallel b\dash$ at 35~kOe
is more evident in the inset of Fig.~\ref{fig:Tc-angle-theta_35kOe}.
For $\Hbm\parallel b\dash$, $\Tco(\theta)$ is nearly constant when $\theta\ge 89\deg$,
whereas $\Tco(\theta)$ decreases approximately linearly with $\theta$ for $\Hbm\parallel a$.

These results contradict the prediction of the Ginzburg-Landau (GL) model,
where only the orbital pair-breaking effect is taken into account.
Assuming the GL anisotropic-mass model,
$\Hcc(\theta)$ for $\Hbm\parallel b\dash$ should exhibit a sharp cusp-like peak at $\theta=90\deg$ 
because of the two dimensionality due to the FIDC,
whereas $\Hcc(\theta)$ for $\Hbm\parallel a$ the peak at $\theta=90\deg$ should be bell-shaped.
Thus $\Hcc(\theta)$, therefore $\Tco(\theta)$, 
should decrease more rapidly for $\Hbm\parallel b\dash$ than for $\Hbm\parallel a$
when the field is slightly tilted from the $ab$ plane.
This contradiction to the GL model suggests that we need to take into account the Pauli effect
to understand this behavior in tilted fields, as will be discussed in Sec.~\ref{subsec:tilt-angle}

%%%%%%%%%%%%%%%%%%%%%%%%%%%%%%%%%%%%%%%%%%%%%%%%
%%%%%%%%%%%%%%%%%%%%%%%%%%%%%%%%%%%%%%%%%%%%%%%%

%%%%%%%%%%%%%%%%%%%%%%%%%%%%%%%%%%%%%%%%%%%%%%%%
%%%%%%%%%%%%%%%%%%%%%%%%%%%%%%%%%%%%%%%%%%%%%%%%

\section{Discussion}

\subsection{Orbital-limited regime}

In the vicinity of $H=0$, all the onset curves presented in Fig.~\ref{fig:phase_diagram}
exhibit linear temperature dependences.
The initial slope of the onset curves ${\mathrm d}\Hcc(T)/{\mathrm d}T|_{T=\Tcz}$ 
are $-67$~kOe/K for $\Hbm\parallel a$,
$-36$~kOe/K for $\Hbm\parallel b\dash$,
and $-1.5$~kOe/K for $\Hbm\parallel \cstar$ for Sample \#1.
For Sample \#2, they are
$-57$~kOe/K for $\Hbm\parallel a$,
$-27$~kOe/K for $\Hbm\parallel b\dash$,
and $-1.5$~kOe/K for $\Hbm\parallel \cstar$.

These linear dependences can be analyzed in a GL model for a type-II superconductor.
First, from the initial slope of $\Hcc(T)$ at $T=\Tcz$, 
we estimate the orbital pair-breaking field $\Horb$ 
using the so-called Werthamer-Helfand-Hohenberg formula
for a clean limit~\cite{Helfand1966PhysRev},
\begin{align}
\Horb = -0.727 \left.\frac{{\mathrm d}\Hcc(T)}{{\mathrm d}T}\right|_{T=\Tcz}\Tcz.
\end{align}
From this equation, we obtain $\Horb$ for Sample \#1 as 70~kOe for $\Hbm\parallel a$,
38~kOe for $\Hbm\parallel b\dash$, and 1.5~kOe for $\Hbm\parallel \cstar$.
For Sample \#2, we also obtain 
60~kOe for $\Hbm\parallel a$,
28~kOe for $\Hbm\parallel b\dash$, and
1.5~kOe for $\Hbm\parallel \cstar$.
It should be noted that $\Horb$ for $\Hbm\parallel a$ is larger than the experimental field range in this study.
Therefore, at temperatures well below $\Tcz$, the orbital depairing effect is 
insignificant for $\Hbm\parallel a$ in the present field range.
%Hc2_orb; Sample #1
%H//a: 70.3107901405002 kOe
%H//b': 37.8292846987125 kOe
%H//c*: 1.5830738316639 kOe
%Hc2_orb; Sample #2
%H//a: 59.5794122286558 kOe
%H//b': 28.1340892344725 kOe
%H//c*: 1.52154600559803 kOe    

Next, we evaluate the GL coherence lengths $\xi(T)$ at $T=0$
using a relation 
\begin{align}
H\spsm{orb}_{{\mathrm c2}\,\parallel\,i}=
\frac{\Phi_0}{2\pi} \frac{1}{\xi_j(0)\xi_k(0)}\,,
\end{align}
where $\Phi_0$ is the flux quantum, and $(i,j,k)$ correspond to $(a,b\dash,\cstar)$
and to their cyclic permutations.
From the initial slopes
we obtain $\xi_a(0)=450$~\AA, $\xi_{b\dash}(0)=240$~\AA, and $\xi_{\cstar}(0)=10$~\AA\ for Sample \#1,
and $\xi_a(0)=490$~\AA, $\xi_{b\dash}(0)=230$~\AA, and $\xi_{\cstar}(0)=13$~\AA\ for Sample \#2.
These values agrees well with the previous report for \tmc~\cite{Murata1987}.
These coherence lengths are much shorter than the mean-free paths $l$ in Table~\ref{tab:samples}.
We therefore again confirm that both samples are in a clean limit.
Note that $\xi_{\cstar}(0)$ is comparable to the inter-layer distance, 
which is approximately 13~\AA.

We can also estimate the transfer integral $t$ of each direction
within a GL theory with a orthorhombic tight-binding model~\cite{Gorkov1985}.
The transfer integrals are related to the $\Hcc$ near $H=0$ through the equations
\begin{align}
H_{{\mathrm c2}\,\parallel\,a} (T)& = \frac{\Tcz - T}{\Tcz} 
\frac{6\pi^2\tilde{c}\hbar\Tcz^2}{7\zeta(3) e \tb  \tc b\dash \cstar}\ \ ,
\label{eq:gl-a}\\
H_{{\mathrm c2}\,\parallel\,b\dash} (T)& = \frac{\Tcz - T}{\Tcz} 
\frac{12\pi^2\tilde{c}\hbar\Tcz^2}{7\zeta(3) e \tc \ta  \cstar a}\ \ ,
\label{eq:gl-b}\\
H_{{\mathrm c2}\,\parallel\,\cstar} (T)& = \frac{\Tcz - T}{\Tcz} 
\frac{12\pi^2\tilde{c}\hbar\Tcz^2}{7\zeta(3) e \ta  \tb  ab\dash}\ \ ,
\label{eq:gl-c}
\end{align}
where $\tilde{c}$ is the speed of light, $\zeta(x)$ is the Riemann's zeta function, 
and $a/2=3.542$~\AA, $b\dash=7.158$~\AA, and $\cstar=13.119$~\AA\ are 
the orthorhombic inter-site distances 
evaluated from the crystal structure at 7~K~\cite{Pevelen2001EurPhysJB}
In order to take into account the fact that for an anisotropic superconductivity
$\langle|\Psi|^2\rangle=1/2$, which is the average over the Fermi surface
of the normalized superconducting order parameter,
we should multiply the right hand side of Eqs.~\eqref{eq:gl-a}--\eqref{eq:gl-c} by 2.
As a result, we obtain practical equations between the transfer integrals and
the slope ${\mathrm d}\Hcc(T)/{\mathrm d}T$ at $T=\Tcz$,
\begin{align}
\left.\frac{{\mathrm d}H_{{\mathrm c2}\,\parallel\,a}(T)}{{\mathrm d}T}\right|_{T=\Tcz}
& = -\frac{98.7\times 10^{3}}{\tb  \tc}\Tcz,
\label{eq:gl-a2}\\
\left.\frac{{\mathrm d}H_{{\mathrm c2}\,\parallel\,b\dash}(T)}{{\mathrm d}T}\right|_{T=\Tcz} 
& = -\frac{199\times 10^{3}}{\tc \ta }\Tcz,
\label{eq:gl-b2}\\
\left.\frac{{\mathrm d}H_{{\mathrm c2}\,\parallel\,\cstar}(T)}{{\mathrm d}T}\right|_{T=\Tcz}
& = -\frac{365\times 10^{3}}{\ta  \tb }\Tcz,
\label{eq:gl-c2}
\end{align}
where $\Hcc$ is in a unit of kOe and $\Tc$ and the transfer integrals are in K.
Applying these equations to our phase diagrams in Fig.~\ref{fig:phase_diagram}, 
we estimated the transfer integrals
to be $\ta =1200$~K, $\tb=310$~K, and $\tc=7.0$~K for Sample \#1,
whereas $\ta =1200$~K, $\tb=290$~K, and $\tc=8.5$~K for Sample \#2.
These values coincide with each other except for $\tc$ and 
they agree reasonably with the realistic band parameters~\cite{IshiguroYamajiText}.
This agreement of the band parameters supports 
the assumption that $\Hcc(T)$ is governed by the orbital limitation
at low fields in all three directions. 
We note that there is a relatively large difference between $\tc$ of Sample \#1 and Sample \#2.
Although the quality of Sample \#2 is poorer, 
Sample \#1 has a smaller $\tc$.
This smaller $\tc$ for Sample \#1 probably indicates that carriers in Sample \#1 are 
strongly confined in the $ab$ plane
due to the onset of the FIDC for \Hvec${}\parallel b\dash$ even in a small field.
The effect of the FIDC on superconductivity will be discussed in the next subsection.

The values obtained from the above analysis are summarized also in Table~\ref{tab:samples}.

\subsection{Influence of the field-induced dimensional crossover}\label{subsec:FIDC}

As we have briefly reviewed in Sec.~\ref{sec:Introduction}, 
the FIDC should strongly suppress the orbital pair-breaking effect~\cite{Lebed1986},
leading to a stabilization of superconductivity in higher fields than $\Horb$.
The apparent absence of the orbital limit in high fields for $\Hbm\parallel b\dash$ for Sample \#1 
agrees with this expectation.
In addition, the effect of the FIDC should also affect the angle dependence of $\Tco(\phi)$.
Indeed, we observed dips in the $\Tco$ curves of Sample \#1
at $|\phi|=\pm\phidip=\pm 17\deg$, as shown in Fig.~\ref{fig:Tc-angle}.
This angle of the dips is very close to the onset angle of the FIDC $\phidco=(19\pm 1)\deg$,
which we defined as the angle above which
$\Rzz(T)$ in the normal state above 20~kOe exhibits non-metallic behavior for Sample \#1.
This onset angle essentially corresponds to the angle above which the closed and semi-closed orbits
of the semi-classical motion of carriers on the Fermi surface disappear~\cite{Yoshino1999JPhysSocJpn}.

In the previous report, we have attributed these dips 
to the interplay between the orbital pair-breaking effect and the FIDC~\cite{Yonezawa2008.PhysRevLett.100.117002}.
Assuming the absence of the FIDC, $\Tco(\phi)$ should exhibit the largest value 
for \Hvec\ parallel to the most-conductive $a$ direction, i.\,e. $\phi=0\deg$,
and $\Tco(\phi)$ would gradually decrease as $|\phi|$ increases.
In reality, however, if the field is rotated from the $a$ axis to the $b\dash$ axis,
the dimensionality start to be lowered by the onset of the FIDC,
leading to an enhancement of superconductivity for $|\phi|>\phidco$.
This competition between the suppression of $\Tco(\phi)$ due to the usual orbital effect
and the enhancement of $\Tco(\phi)$ due to the onset of the FIDC
results in the dips of $\Tco(\phi)$ around $|\phi|=\phidco$.

%However, to the best of knowledge, direct evidence of this effect 
%of the FIDC on superconductivity has not been reported so far.
Comparison of the results of the two samples with different magnitudes 
of the FIDC provides us with a more convincing evidence of the effect of the FIDC to the superconductivity.
First, the large difference in the onset curves in the $H$-$T$ phase diagram 
observed \textit{only} for $\Hbm\parallel b\dash$ is attributable to the difference of the magnitude of the FIDC.
Secondly, in Fig.~\ref{fig:Tc-angle} we compare the $\Tco(\phi)$ curves of Sample \#1 with those of Sample \#2.
It is clear that the curves for Sample \#1 and the curves for Sample \#2 nearly coincide with each other
when $|\phi|$ is smaller than $\phidip$.
%This fact indicates that for $|\phi|<\phidco$, where the electronic state is essentially 3D,
%the stability of superconductivity of Sample \#1 and that of Sample \#2 is nearly the same.
In contrast, the two curves strongly deviate from each other for $|\phi|>\phidip$:
For Sample \#1, which exhibits a strong FIDC, 
there is an enhancement of $\Tco(\phi)$ for $|\phi|>\phidco$;
whereas $\Tco(\phi)$ for Sample \#2, in which FIDC is much weaker,
does not exhibit enhancement and keeps decreasing with increasing $|\phi|$.
Thus these differences of the $H$-$T$ and $T$-$\phi$ phase diagrams
observed only in the 2D angle range ($|\phi|>\phidco$)
strongly supports the scenario that the FIDC indeed enhances $\Tco$ in this angle range.

\subsection{Pauli-limited regime}

Here we first estimate an appropriate value of $\Hp$ using experimental results.
Reminding the fact that the SC condensation energy $U\subm{c}$ 
is equal to the energy gain of the Zeeman splitting at $\Hp$,
one obtains the relation
\begin{align}
U\subm{c}=\frac{H\subm{c}^2}{8\pi}V\subm{m}=\frac{1}{2}(\chi\subm{n}-\chi\subm{sc})\Hp^2,
\label{eq:Pauli1}
\end{align}
where $H\subm{c}$ is the thermodynamic critical field of superconductivity,
$V\subm{m}$ is the volume of one-mole \tmc,
and $\chi\subm{n}$ and $\chi\subm{sc}$ are the spin susceptibility in the normal state 
and in the SC state, respectively~\cite{Clogston1962}.
Assuming $\chi\subm{sc}=0$ for a singlet superconductor, 
Eq.~\eqref{eq:Pauli1} can be transformed into
\begin{align}
\Hp = H\subm{c}\sqrt{\frac{V\subm{m}}{4\pi\chi\subm{n}}} .
\end{align}
We refer to the experimental values $H\subm{c}=44$~Oe 
obtained from the integration of the electronic specific heat 
in the SC state by Brussetti \etal~\cite{Brusetti1983JPhysC}
and $V\subm{m}=665$~\AA$^3$ from the crystal-structure study by P\'{e}velen \etal~\cite{Pevelen2001EurPhysJB}
The value of $\chi\subm{n}$ at low temperatures ranges over $260$--$340\times 10^{-6}$~emu/(mol-\tmc)
depending on literatures~\cite{Miljak1983JPhys,Miljak1985MolCrystLiqCryst,Haddon1994},
mainly due to the difficulty in the subtraction of the large diamagnetic contribution of the ionic cores.
Thus the estimated value of $\Hp$ ranges from 23 to 26~kOe,
in agreement with the simple estimation $\Hp \sim 26.7$~kOe 
from the relation for an isotropic gap, $\Hp/\Tcz=18.4$~kOe/K~\cite{Clogston1962}, 
used in our previous report~\cite{Yonezawa2008.PhysRevLett.100.117002}.

As discussed in Sec.~\ref{subsec:phase_diagram},
the in-plane field-angle dependence of $\Tco$ becomes asymmetric with respect to 
the $a$ and $b\dash$ axes above 25--30~kOe.
From the fact that this field agrees with the estimated $\Hp\sim23$--26~kOe,
we infer that this asymmetry is attributable to the Pauli pair-breaking effect.

\begin{figure}
\begin{center}
\includegraphics[width=7.5cm]{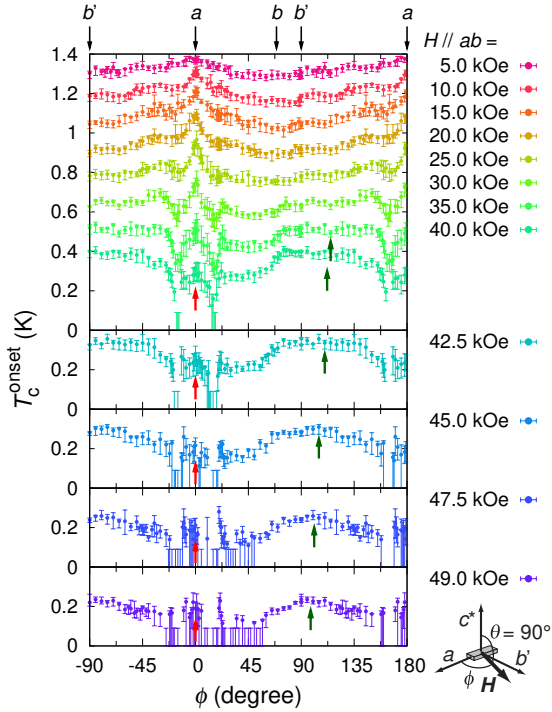}
\caption{(color online)
Detailed data of $\Tco(\phi)$ for Sample \#1, plotted in the range $-90\deg \le \phi \le 180\deg$.
The curves in the top panel
correspond to the data for $H=5$, 10, $\dots$ , and 40~kOe (5-kOe steps from top to bottom), respectively.
The lower four panels present the data for $H=42.5$, 45.0, 47.5, and 49.0~kOe, respectively.
The points for $\phi>90\deg$ is from the same data as the points for $\phi<0\deg$.
The small arrows indicate the two directions of maxima of $\Tco(\phi)$ in high-field regime, 
the $a$ axis (red arrows at $\phi=0\deg$) and the \x\ axis (dark green arrows in the range $\phi=95$--$110\deg$).
\label{fig:Tc-angle_-90-180}}
\end{center}
\end{figure}

In order to understand the meaning of the asymmetry more clearly, 
we plotted $\Tco(\phi)$ of Sample \#1 in Fig.~\ref{fig:Tc-angle_-90-180}
in the range $-90\deg \le \phi \le 180\deg$.
In this figure the points used for $\phi > 90\deg$ are the same data as the points for $\phi<0\deg$.
This two-fold symmetry of the data within a plane is required 
due to the inversion symmetry of the crystal structure, 
which belongs to the $P_{\bar{1}}$ space group 
regardless of the anion ordering~\cite{Pevelen2001EurPhysJB}.
In low fields $\Tco(\phi)$ exhibits a sharp maximum at $\phi=0\deg$ ($\Hbm\parallel a$)
and a broad minimum at $\phi=90\deg$ ($\Hbm\parallel b\dash$).
However, it is clear that above $\Hp$ the $b\dash$ axis is no longer the direction of a minimum of $\Tco(\phi)$,
and a broad peak of $\Tco(\phi)$ starts to emerge centered around $\phi=110\deg$ (or equally $\phi=-70\deg$).
What is more, this peak position apparently moves toward $\phi=90\deg$ as field increases,
as indicated by the green arrows in Fig.~\ref{fig:Tc-angle_-90-180}.
At 49~kOe, the largest field available in this study of the angle dependence, 
the peak position becomes approximately $\phi\sim98\deg$.
This new peak corresponds to the new principal axis \x\ of $\Tco(\phi)$ 
in our previous report~\cite{Yonezawa2008.PhysRevLett.100.117002}.
Thus the asymmetry shown in Fig.~\ref{fig:asymmetry} seems intimately related to 
the appearance of the second peak in $\Tco(\phi)$.

On the other hand, the other peak at $\phi=0\deg$ exists even above $\Hp$ up to 49~kOe.
Therefore, the SC state above $\Hp$ for Sample \#1 can be 
divided into two states separated by the dips of $\Tco(\phi)$.
One is the SC state around the $a$ axis,
where the electronic state is anisotropic 3D,
with the highest $\Tco(\phi)$ at $\phi=0\deg$.
The other is the SC state in $|\phi|>\phidco$, 
the angle range of the reduced dimensionality due to the onset of the FIDC,
with the maximum of $\Tco(\phi)$ for $\Hbm\parallel \Xbm$.
These two states have different positions of maximum of $\Tco(\phi)$, and
the dimensionality of the electronic state is also different.
Therefore, a separate discussion should be necessary for these two regimes.

\subsubsection{Pauli-limited regime in the angle range of the reduced dimensionality}\label{subsec:Pauli_Hb}

The high-field superconducting state for $|\phi|>\phidco$ 
is the most stable for fields parallel to the \textit{characteristic direction} $\Xbm$,
which is different from both the crystalline $b\dash$ and $b$ axes.
The appearance of this ``incommensurate'' characteristic direction 
strongly suggests that the SC state for $|\phi|>\phidco$ above $\Hp$
has an additional spatial symmetry breaking compared to the low-field SC state.

%%%%%%%%%%%%%%%%%%%%%%%%%%%%%%%%%%%%%%%%%%%%%%%%
%%%%%%%%%%%%%%%%%%%%%%%%%%%%%%%%%%%%%%%%%%%%%%%%

Now we discuss the origin of the new principal axis \x,
emerging in the high-field superconducting state.
Possible candidates for such a superconductivity with a characteristic direction 
include an FFLO state for a singlet superconductivity and a spin-triplet superconductivity
with Cooper pair spins polarized along a certain direction.

\begin{figure}
\begin{center}
\includegraphics[width=7.5cm]{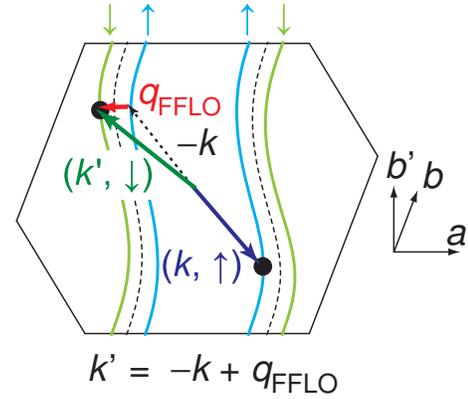}
\caption{(color online)
Schematic drawing of a pairing combination [$(\kbm,\,\uparrow)$, $(-\kbm+\qFFLO,\,\downarrow)$] 
in an FFLO state for a Q1D system. 
The curves labeled with $\uparrow$ and with $\downarrow$ are the spin-up Fermi surfaces
and the spin-down Fermi surfaces, respectively,
and the broken curve is the Fermi surface at $H=0$.
Black circles represent the positions of carriers forming a pair with $\qFFLO$ in the $k$-space.
Note that $\qFFLO$ is nearly equal to the nesting vector between the spin-up and the spin-down Fermi surfaces $\qnest$.
\label{fig:FFLO}}
\end{center}
\end{figure}

An FFLO state is a superconducting state with pairing combinations 
[$(\kbm,\,\uparrow)$, $(-\kbm+\qFFLO,\,\downarrow)$]
instead of the ordinary Cooper pairs [$(\kbm,\,\uparrow)$, $(-\kbm,\,\downarrow)$].
If the Pauli pair-breaking effect is dominant in a singlet superconductor, 
the ordinary Cooper pairs are destroyed at $\Hp$.
However, in an FFLO state, 
pairs can be formed on some part of the split Fermi surfaces,
gaining both of the Zeeman energy and some portion of the condensation energy.
In this case, the $(\kbm,\,\uparrow)$ quasiparticle cannot find its ``partner'' of $(-\kbm,\,\downarrow)$ 
because $-\kbm$ is not located on the spin-down Fermi surface as shown in Fig.~\ref{fig:FFLO}.
Therefore, the $(\kbm,\,\uparrow)$ quasiparticle forms a pairing combination with the quasiparticle of
$(\kbm\dash,\,\downarrow)$, where $\kbm\dash$ is slightly shifted from $-\kbm$: $\kbm\dash=-\kbm+\qFFLO$.
It is worth noting that $\qFFLO$ should nearly match 
\textit{the nesting vector between the spin-up and the spin-down Fermi surfaces} $\qnest$
in case of a strong nesting between the split Fermi surfaces~\cite{Shimahara1994.JPhysSocJpn.50.12760,Shimahara1997.JPhysSocJpn.66.541}.
This is because a large number of quasiparticles can find their ``partners'' and 
form pairs of [$(\kbm,\,\uparrow)$, $(-\kbm+\qFFLO,\,\downarrow)$] 
and thus the gain of the condensation energy also becomes large 
if there is a strong nesting and $\qFFLO\sim\qnest$.

A Q1D system is a typical example of a system with a strong nesting
between the spin-up and the spin-down Fermi surfaces,
whose $\qnest$ is nearly perpendicular to its Fermi surface sheet
and thus nearly parallel to the most conductive direction.
Because of this nesting nature, an FFLO state in a Q1D system is greatly stabilized,
whose $\Tc$ depends on the applied magnetic field as $\Tc\propto 1/H$ 
assuming the absence of the orbital effect~\cite{Machida1984PhysRevB}.
This means that $\Tc$ may remain finite up to very high magnetic fields.

%%%%%%%%%%%%%%%%%%%%%%%%%%%%%%%%%%%%%%%%%%%%%%%%
%%%%%%%%%%%%%%%%%%%%%%%%%%%%%%%%%%%%%%%%%%%%%%%%

For \tmx\ in $\Hbm\parallel b\dash$, it has been discussed using orthorhombic band structures 
that an FFLO state with $\qFFLO\parallel a$ 
becomes stable with a help of the FIDC~\cite{Lebed1986,Dupuis1994,Miyazaki1999}.
Although we are not aware of a publication on the in-plane field-angle variation of this FFLO state,
we expect that for $|\phi|>\phidco$, where the conductive planes are decoupled,
this FFLO state is still stable.
However, the direction of $\qFFLO$, matching the nesting vector, should be tilted from the $a$ axis,
because of the triclinic Fermi surface of \tmx.
In addition, $\qFFLO$ may vary with increasing the field 
because the separation between the spin-up and the spin-down Fermi surfaces depends on $|\Hbm|$.
Within this scenario, one possible explanation of $\Xbm$, 
is that $\Xbm$ corresponds to the field direction which favors the spatial structure of the FFLO state,
which should be closely related to $\qFFLO$.
Because the direction of $\qFFLO$ is expected to depend on the field strength as we explained,
the spatial structure of the FFLO state and thus the direction of $\Xbm$ may also rotate in increasing field, 
which is consistent with our experimental results.
Another possible explanation, 
in which we assume the field-strength dependence of the direction of $\qFFLO$ 
is too small to be attributable to the observed rotation of $\Xbm$, 
is a competition between the FIDC, which favors $\Hbm\parallel b\dash$, and the FFLO state which favors 
a field direction favorable to the spatial structure of the FFLO state.

On the other hand, if \tmc\ is a triplet superconductor,
polarized Cooper pair spins may cause the additional spatial symmetry breaking of superconductivity.
Assuming that the spin of Cooper pairs is fixed to one direction,
superconductivity is not affected by a Pauli effect when field is exactly parallel to the spin
while it is suppressed in other field directions.
%Such an anisotropy may appear above $\Hp$, where Pauli pair-breaking effect become more important 
%than orbital pair-breaking effect.
Therefore, the direction of the new principle axis should correspond to 
the direction of the polarized spin of the Cooper pairs.
In this case, however, it seems difficult to explain 
why the ``incommensurate'' direction of $\Xbm$ is chosen
and why the direction of $\Xbm$ rotates as the field increases.
%In this scenario, probably a strong spin-orbit interaction should be exist.
In addition, as will be discussed in the next subsubsection, 
the high-field phase for $\Hbm\parallel a$ is more sensitively suppressed by impurity scattering
than the low-field superconductivity.
This supports the FFLO scenario but probably not the triplet scenario.

We note that strong contribution of SC fluctuations seems to exist in \tmc\ in high field regime. 
The existence of SC fluctuations is evident from the absence of zero resistance
despite a bulk superconductivity confirmed through the torque measurement~\cite{Oh2004}.
The broadening of the SC transition between $\Tco$ and $\Tc^{R=0}$ in high fields 
shown in Fig.~\ref{fig:phase_diagram}
also supports the contribution of fluctuations.
Adachi and Ikeda~\cite{Adachi2003} suggested that the transition from the normal state to the FFLO state
should be rounded into a crossover in the presence of strong SC fluctuations.
The absence of the observation of a clear phase transition to the FFLO state
may be attributed to this effect of SC fluctuations.

For this angle range, comparison between the influence of the Pauli effect
of Sample \#1 and Sample \#2 is not straightforward,
because the difference of the scattering time contributes mainly through the FIDC.
It is not easy to judge whether the absence of the upturn of the onset curve for Sample \#2 
originates from the FIDC or from the stability of superconductivity against impurity scatterings.
However for $\Hbm\parallel a$, where the FIDC is absent, comparison between the two samples is meaningful,
because the difference of the onset curve should be related only to the stability of superconductivity.
This comparison will be discussed in the next subsubsection.

\subsubsection{Pauli-limited regime for the 3D angle range}

For the angle range $|\phi|<\phidco$, the electronic system is 3D and
the $a$ axis remains the principal axis up to the highest field.
Therefore, the above discussion for $|\phi|>\phidco$ cannot be applied for this angle range.

For $\Hbm\parallel a$, both onset curves in Fig.~\ref{fig:phase_diagram} 
for Sample \#1 and \#2 exhibit saturating behavior above 20~kOe.
The saturating behavior indicates that the onset curves are mainly governed by the Pauli pair-breaking effect 
in the high-field regime for $\Hbm\parallel a$.
The insignificance of the orbital effect is also clear from 
$\Horb$ of 70~kOe for $\Hbm\parallel a$, being much larger than 20~kOe.

The two curves for Sample \#1 and \#2 nearly coincide below 40~kOe.
However, the upturn of the onset curve above 40~kOe is observed \textit{only} in Sample \#1.
In contrast to the case of $\Hbm\parallel b\dash$, 
this deviation of the onset curves is not attributable to FIDC.
Therefore, the difference between the two curves
directly arises from the difference of the stability of superconductivity against impurity scatterings:
The superconducting state above 40~kOe is more sensitively destroyed by impurities
than the superconductivity in the low-field regime.
Reminding the prediction that the FFLO state is easily suppressed by impurity scatterings~\cite{Gruenberg1966}
and the apparent Pauli-limited behavior of the $H$-$T$ phase diagram,
one possible explanation of the superconducting phase above 40~kOe for $\Hbm\parallel a$ is an FFLO state.
This scenario is consistent with the discussion of the FFLO scenario
for $|\phi|>\phidco$ in the previous sub-subsection.

To our knowledge, there are only a few studies of FFLO states of Q1D systems 
in magnetic fields parallel to the most conducting direction.
A theoretical work on the superconductivity under a strong ferromagnetic molecular field~\cite{Machida1984PhysRevB}
with a neglected orbital effect might be in a similar situation as \tmc\ for $\Hbm\parallel a$.
This theory predicted a phase diagram that resembles the onset curve of Sample \#1 
for $\Hbm\parallel a$ in Fig.~\ref{fig:phase_diagram}(a).
An FFLO state proposed for doped two-leg ladder cuprates 
using a $t$-$J$ model~\cite{Roux2006} might also provide some explanation,
although a theory adapted to \tmc, a coupled chain system, needs to be developed.
In this model, an FFLO state is realized with  antiferromagnetic fluctuations.
Theories more appropriate for \tmc\ in $\Hbm\parallel a$ are required.

We should comment here on the field range $\Hp \lesssim H \lesssim 40$~kOe.
The value $\Hp=23$--26~kOe may apparently differ from the field range where
the discrepancy between the onset curve for $\Hbm\parallel a$ 
of Sample \#1 and that of Sample \#2 occurs, which is about 40~kOe.
Nevertheless, it should be noted that the suppression of $\Tc$ by a small tilt of the magnetic field
becomes significant above $\Hp$, as shown in Sec.~\ref{subsec:phase_diagram_tilt}.
One possible interpretation of this strong suppression is 
that another FFLO state, which is sensitively destroyed by a small tilt of the magnetic field
while relatively stable against impurity scatterings, 
is realized between $H\sim \Hp$ and $H\sim 40$~kOe,
although there is no clear experimental evidence for this speculation.
We elaborate on this interpretation again in the next subsection.

On the other hand, if the superconducting phase for $|\phi|>\phidco$ is a triplet state,
the high-field phase for $\Hbm\parallel a$ should also be a triplet state.
In this case, it seems difficult to explain the observed difference 
between the stability against impurity scatterings of the high-field state and of the low-field state.
Thus so far the FFLO scenario is more plausible than the triplet scenario.

\subsection{Effect of tilting the magnetic field}\label{subsec:tilt-angle}

As shown in Figs.~\ref{fig:Tc-angle_87deg} and \ref{fig:Tc-angle-theta_35kOe} 
the suppression of $\Tco$ due to the tilt of the magnetic field out of the $ab$ plane 
differs between the ranges $|\phi|<\phidco$ and $|\phi|>\phidco$.
We have discussed in Sec.~\ref{subsec:phase_diagram_tilt} 
that the $\theta$ dependence of $\Tco$ contradicts the simple GL model.
More detailed calculation by Vaccarella and S{\'{a}} de Melo~\cite{Vaccarella2001.PhysRevB.64.212504}
for an equal-spin triplet scenario, where the Pauli pair-breaking is absent,
also predicted a steep reduction of $\Tc$ near $\theta=90\deg$ for $\Hbm$ in the $b\dash\cstar$ plane,
being different from the observed weak $\theta$ dependence for $\phi=90\deg$.

On the contrary, $\theta$ dependence of $\Tco$ should be rather isotropic 
if it is mainly governed by the Pauli pair-breaking effect,
which should be nearly isotropic in \tmx, where the spin-orbit coupling is not very strong.
Therefore, the weak $\theta$ dependence of $\Tco$ for $\phi=90\deg$ ($\Hbm\parallel b\dash\cstar$)
suggests a dominance of the Pauli pair-breaking effect in the determination of $\Tco$,
although the $H$-$T$ phase diagram for $\Hbm\parallel b\dash$ in Fig.~\ref{fig:phase_diagram} 
does not appear to be Pauli-limited.

In the case of an FFLO scenario,
an interplay between a modulation of the order parameter in a FFLO state and the structure of the vortices
should be taken into account in order to fully understand the effect of a tilt of the magnetic field.
For example, in a purely 2D system in high fields above $\Hp$, FFLO states without vortices emerge 
if the field is exactly parallel to the conductive plane.
A small tilt of the magnetic field out of the conductive plane
induces vortices in the plane and 
leads to a coexistence of the vortex lattice and the spatial modulation of the order parameter.
Such states are interpreted as vortex states with large Landau quantum number
$n\ge 1$~\cite{Yang2004PhysRevB},
whereas $n$ of the ordinary vortex state is zero~\cite{TinkhamText}.
It has been argued that a large-$n$ vortex state can emerge at low temperatures 
if the tilt angle of the field from the plane is quite small,
while for a large tilt angle only the ordinary vortex lattice with $n=0$ can be realized
in the whole temperature and field region~\cite{Bulaevskii1974JETP,Shimahara1997.JPhysSocJpn.66.3591,Klein2004PhysRevB}.
According to Shimahara and Rainer~\cite{Shimahara1997.JPhysSocJpn.66.3591}, 
if the tilt angle is small enough,
the $\Hcc(T)$ curve is indistinguishable from the curve for the parallel field.
In other words, the suppression of $\Tco$ due to the tilt of the field can be small for a small tilt angle.
The observed weak $\theta$ dependence of $\Tco$ for $|\phi|>\phidco$ 
might be consistent with their prediction,
although theories more appropriate for \tmc\ should be developed.

The rapid reduction of $\Tco$ in slightly-tilted fields for $H>\Hp$ and $|\phi|<\phidco$ is rather unusual.
Within the FFLO scenario discussed in the previous subsection, 
one possible explanation of the sensitivity against the tilt of the magnetic field
is that the 3D nature of the electronic system and/or the chain-like crystal structure along the $a$ axis
might enhance the competition between the vortex lattice and 
the spatial order parameter structure with $\qFFLO$, which is expected to be nearly parallel to the $a$ axis.
The SC state at $\theta=87\deg$ and $\phi=0\deg$ is probably in the ordinary vortex state with $n=0$
and completely depaired by the Pauli effect.

\section{Conclusion}

We have studied the in-plane anisotropy of the SC onset temperature $\Tco$ of \tmc\
and we compared the results for two samples with almost the same onset temperatures but 
different relaxation time of the quasiparticles.
At lower fields, the linear temperature dependence of 
the onset curves in the $H$-$T$ phase diagrams for $\Hbm\parallel a$, 
$\Hbm\parallel b\dash$, and $\Hbm\parallel \cstar$ can be analysed within an anisotropic 3D GL model,
indicating the dominance of the orbital pair-breaking effect.
The dips of $\Tco(\phi)$ at $\phi=\pm\phidip=\pm17\deg$ observed only for Sample \#1 above 20~kOe
as well as the results of the GL analysis 
are attributed to the stabilization of superconductivity due to the field-induced dimensional crossover (FIDC).
The unusual upturn of the onset curve in the $H$-$T$ phase diagram
was observed for $\Hbm\parallel b\dash$ as well as $\Hbm\parallel a$, but only for Sample \#1.
The upturn of the onset curve for $\Hbm\parallel a$ 
suggests the existence of another superconducting state
that overcomes the Pauli pair-breaking effect but is easily suppressed by impurity scatterings.
Meanwhile, the upturn for $\Hbm\parallel b\dash$ results from the combination of the effect of the FIDC
and the occurrence of another superconducting state with a new principal axis \x\ of $\Tco(\phi)$.
This high-field state for $\Hbm\parallel b\dash$ has an additional spatial symmetry breaking
and has an ``incommensurate'' characteristic direction.
We discussed that the most favorable scenario so far to explain these observations
is that \tmc\ is a singlet superconductor and high-field phases are FFLO states,
by all the present experimental results being taken into account.
The suppression of $\Tco$ by a tilt of the applied field from the conductive plane
does not favor triplet scenarios either.
The difference of the suppression of $\Tco(\theta)$ between 
for $\Hbm\parallel a\cstar$ and for $\Hbm\parallel b\dash\cstar$
might be related to the competition of the spatial structure
of the FFLO order parameter and the structure of vortex lattice.
We note that the recent NMR study, which reported that the density of states at the Fermi level 
recovers to the normal state value in the SC phase above 20~kOe for 
both $\Hbm\parallel a$ and $\Hbm\parallel b\dash$~\cite{Shinagawa2007},
would support the FFLO scenario.

In conclusion,
we have revealed that \tmc\ is a quite interesting material
in which we can observe the unconventional interplay and competition among many effects on superconductivity,
such as the orbital pair-breaking effect, the Pauli pair-breaking effect, 
the dimensionality of the electronic system, 
and possible emergence of FFLO states and 
its response to impurity scatterings and to the tilt of the magnetic field.
%In addition, this interplay can be varied with changing the magnetic field strength and direction.
Because its electronic structure is typical for Q1D systems, 
these anomalous behavior of superconductivity observed in \tmc\ 
are probably universal features in Q1D superconductors.
We believe that further studies of superconductivity in \tmc\ will lead to deeper understandings of 
superconducting phenomena in low dimensional systems.

%\begin{acknowledgements}
%\acknowledgements
\section*{Acknowledgment}
We gratefully acknowledge K. Bechgaard for providing us with good \tmc\ crystals.
We also acknowledge Y. Machida, N. Joo, and M. Kriener for their supports,
and R. Ikeda, D. Agterberg, D. Poilblanc, G. Montambaux, N. Dupuis, 
Y. Suzumura, and A. Kobayashi for useful discussions.
This work has been supported by a Grant-in-Aid for the 21st Century COE 
``Center for Diversity and Universality in Physics'' from  Ministry of Education, Culture, Sports, Science and 
Technology (MEXT) of Japan.
It has also been supported by Grants-in-Aids for Scientific Research from MEXT 
and from Japan Society for the Promotion of Science (JSPS).
One of the authors (S.~Y.) is financially supported as a JSPS Research Fellow.

%\end{acknowledgement}

\bibliography{D:/cygwin/home/Owner/SSP/paper/string,%
D:/cygwin/home/Owner/SSP/paper/TMTSF,%
D:/cygwin/home/Owner/SSP/paper/FFLO,%
D:/cygwin/home/Owner/SSP/paper/SC,%
D:/cygwin/home/Owner/SSP/paper/organic_SC,%
D:/cygwin/home/Owner/SSP/paper/high-Tc,%
D:/cygwin/home/Owner/SSP/paper/measurement_technique,%
D:/cygwin/home/Owner/SSP/paper/textbook,%
../fullpaper_2007}

\end{document}